\documentclass[aps, prb, twocolumn, a4paper, 10pt, nofootinbib, showpacs,reprint,superscriptaddress]{revtex4-2}

\usepackage[utf8]{inputenc}
\usepackage{amsmath}
\usepackage{amssymb}
\usepackage{amsfonts}
\usepackage{amstext}
\usepackage{amssymb}
\usepackage{amsthm}
\usepackage{bbold}
\usepackage{mathtools}
\usepackage{physics}
\usepackage{graphicx}
\usepackage{dcolumn}
\usepackage{bm}
\usepackage{braket}
\usepackage{color}
\usepackage[english]{babel}
\usepackage{xcolor}
\usepackage{setspace}
\usepackage{diagbox}
\usepackage{placeins}
\usepackage{algorithm}
\usepackage{algpseudocode}
\usepackage[inline]{enumitem}
\usepackage{enumerate}
\usepackage[normalem]{ulem}
\usepackage{tikz}
\usepackage[percent]{overpic}

\DeclareUnicodeCharacter{2212}{\ensuremath{-}}

\allowdisplaybreaks
\advance\parskip 0.4pt

\usepackage{hyperref}
\hypersetup{
	colorlinks=true,
	linkcolor=blue,
	filecolor=blue,
	citecolor = red,      
	urlcolor=cyan,
}
\usepackage{orcidlink}
\begin{document}

\renewcommand{\figureautorefname}{Fig.}
\renewcommand{\sectionautorefname}{SM.}
\renewcommand{\subsectionautorefname}{Appendix}

\renewcommand{\equationautorefname}{Eq.}
\renewcommand{\thesection}{\Roman{section}}
\renewcommand{\thesubsection}{\Roman{subsection}}

\title{
Critical behavior of the thermal phase transition of U(1) lattice gauge systems}

\date{\today}
\author{Greta Sophie Reese \orcidlink{0009-0001-9135-7499}}
\email[Contact author: ]{greta.reese@uni-hamburg.de}
\affiliation{Center for Optical Quantum Technologies, University of Hamburg, 22761 Hamburg, Germany}
\affiliation{The Hamburg Centre for Ultrafast Imaging, Hamburg, Germany}
\author{Ludwig Mathey \orcidlink{0000-0003-4341-7335}}
\affiliation{Center for Optical Quantum Technologies, University of Hamburg, 22761 Hamburg, Germany}
\affiliation{The Hamburg Centre for Ultrafast Imaging, Hamburg, Germany}

\begin{abstract}
We model the phase transition of a superconductor as a U(1) lattice gauge system, and determine its critical behavior. For this, we perform Monte Carlo simulations, treating the order parameter field and the gauge field on equal footing, without additional approximations. As the defining correlation function, we determine the order parameter correlation function including a gauge string, thus achieving a gauge-invariant characterization of the long-range behavior explicitly. We obtain a critical exponent $\beta$ that is consistent with the exponent of the U(1) transition of neutral bosons, i.e. of Bose-Einstein condensation. We further extract the anomalous dimension $\eta$ for our parameter sets which is significantly smaller than the U(1) anomalous dimension. We determine the critical behavior of the heat capacity, which displays a temperature depends consistent with an XY transition. These results clarify the universality class of the phase transition of this system.
\end{abstract}

\maketitle

\subsection{INTRODUCTION}
The critical behavior of a phase transition captures how macroscopic order emerges.
The specific universality class is determined by properties such as what type of degrees of freedom are included, what symmetries the system displays, and its dimensionality. In the following, we focus on the transition to the superconducting state. We model this as a U(1) lattice gauge theory, aimed at unconventional superconductors. The model features a complex order parameter field and a gauge field, the electromagnetic vector potential.\\

Previous studies on gauge field systems, explored the histograms $P(|\psi|^2)$ \cite{Mo2002}, finite size-scaling of the free energy \cite{Mo2002}, the moments of the action \cite{Kragset2006}, the planar structure function \cite{Smorgrav2005}, the gauge field correlators $\mathcal{G}(q)=\expval{A_q A_{-q}}$ \cite{Smiseth2005,Herland2010}, and the helicity modulus \cite{Herland2010}. While these properties distinguish different phases, these studies do not evaluate the nonlocal gauge invariant correlation function directly, as we do below. \\

In this work, we determine the critical behavior of the U(1) lattice gauge theory in three dimensions, in an approximation-free manner,  and treating the order parameter field and the gauge field on equal footing. As we describe below, we determine the defining correlation function of the order parameter including the gauge string explicitly, which implies that the resulting analysis of the long-range behavior is gauge invariant.
Using Monte Carlo simulations, we demonstrate that the inclusion of the gauge field results in a critical behavior that is consistent with U(1) universality class, for the example parameter sets studied here.
In particular, we show that the critical exponent $\beta$ is consistent with the U(1) universality class in the presence of a gauge field. Furthermore, we  estimate the anomalous dimension $\eta$ for our parameter sets which is very small compared to the U(1) universality class. Also, we show that the transition of the heat capacity is consistent with an XY-transition.

\subsection{LATTICE GAUGE MODEL}
\begin{figure}
    \centering
    \includegraphics[width=\linewidth]{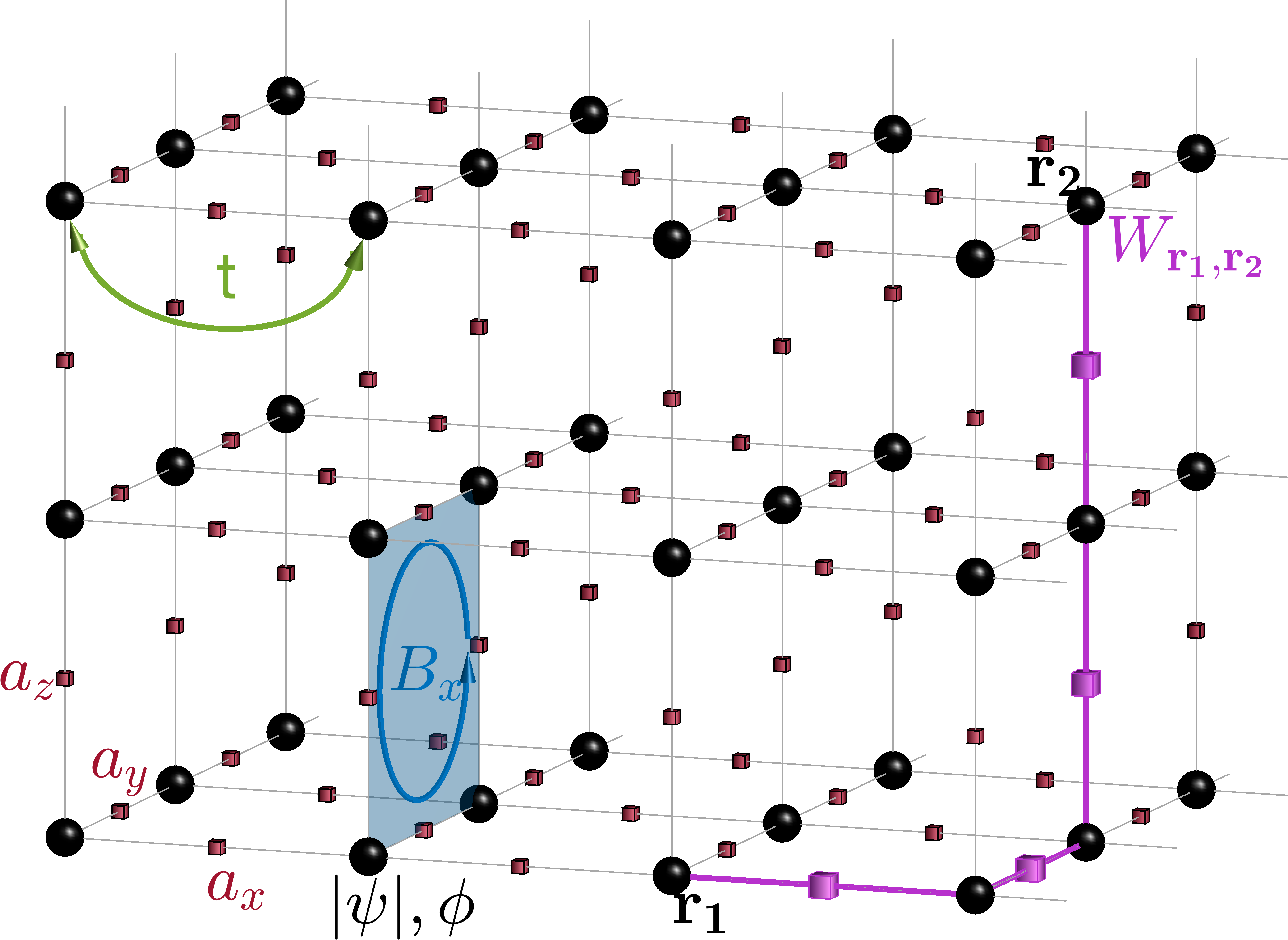}
    \caption{ Sketch of a lattice U(1) gauge theory that we use as a model for the superconducting state.  
    The amplitude $\abs{\psi}$ and the phase $\phi$ of the complex order parameters $\psi$ are defined on the lattice sites, which are represented by black spheres in the sketch. The components $a_j$ of the vector potential, in unitless form, are defined on the links connecting neighboring sites along the directions $j \in {x, y, z}$, depicted as brown cubes. 
    The magnetic $B_i$-fields, which are derived from the surrounding values of the vector potential $a_j$, are defined on the plaquettes of the lattice and are shown exemplary for one $B_x$-value in blue in the sketch. The tunneling strength between the lattice sites is denoted by $t$. For the Wilson line $W_{\boldsymbol{r_1},\boldsymbol{r_2}}$ all vector potential components $a_i$ on the connecting line between $\boldsymbol{r_1}$ and $\boldsymbol{r_2}$ are accumulated (shown in purple).}
    \label{fig1}
\end{figure}

As a model for the superconducting state, we employ a lattice gauge theory in three spatial dimensions to describe the condensation of charged bosons, specifically Cooper pairs with charge $-2e$.  As we describe below, we obtain a low-temperature phase, which displays string order, corresponding to the superconducting state, and a high-temperature phase, which is disordered. In this model, the Cooper pairs are preserved in the high-temperature phase, but not condensed. As such, the criticality of this model applies to strongly-correlated superconducting states featuring pre-formed pairs, but not to conventional superconductors, in which the Cooper pairs break into electrons at the critical temperature.\\
The model is composed of two coupled fields, the order parameter field and the electromagnetic vector potential. The order parameter is represented by a complex scalar field
parameter $\psi = \abs{\psi}\exp{i\phi}$, defined on the sites $\boldsymbol{r}$ of a cubic lattice, i.e. we have field $\psi(\boldsymbol{r})$. The amplitude squared of this field approximately describes the local Cooper pair density. The components of the vector potential $A_j(\boldsymbol{r})$ are defined on the bonds connecting site $\boldsymbol{r}$ to its nearest neighbor in the $j \in \{x, y, z\}$ direction. We represent $A_j(\boldsymbol{r})$ in unitless form via $a_{j,\boldsymbol{r}}=-2edA_{j}(\boldsymbol{r})/\hbar$, where $d$ is the bond length. An illustration of the system is shown in \autoref{fig1}.\\

\begin{figure*}
    \centering
    \begin{overpic}[width=\linewidth]{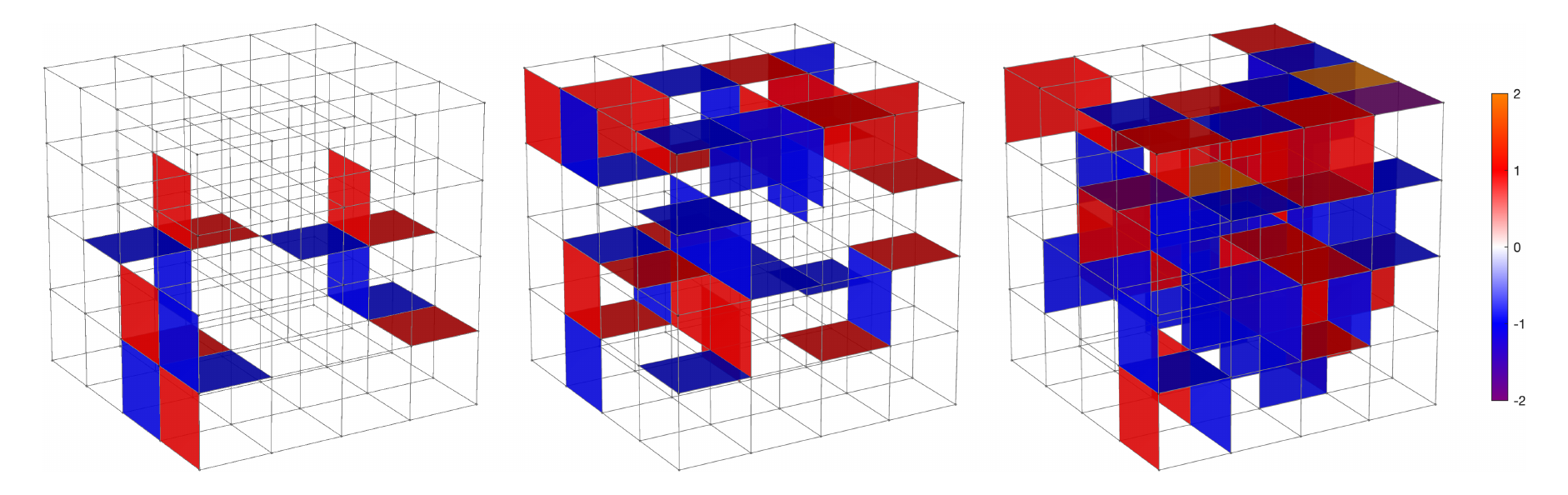}
    \put(5,-1){(a) $\Tilde{T}=0.0253$} 
    \put(40,-1){(b) $\Tilde{T}=0.0272$}
    \put(70,-1){(c) $\Tilde{T}=0.0290$}
    \end{overpic}
    \caption{Snapshots of the vorticity $\nu$ of the system on a lattice of $4^3$ grid points for three typical configurations with increasing temperature. In (a) at $\Tilde{T}=0.0253$ first vortex rings form. (b) At $\Tilde{T}=0.0272$ more complex vortex clusters emerge. In (c) at $\Tilde{T}=0.0290$~$\approx \Tilde{T_c}$ the number of vortices further increases and double vortices form. 
    }
    \label{fig2}
\end{figure*}

To describe the equilibrium properties of the superconducting state and its criticality, we employ a discretized version of the Ginzburg–Landau free energy \cite{Ginzburg1950,Homann2020,Homann2021}, which includes contributions from the local interaction, the kinetic energy of the order parameter, and the electromagnetic field:
\begin{align}
    F=F_{sc}+F_{kin}+F_{em}.
\end{align}
The term $F_{sc}$ captures the contribution from the chemical potential and the local interaction energy, and is given in the lattice representation by
\begin{align}
    F_{sc}
    =\sum_{\boldsymbol{r}} (-\mu |\psi_{\boldsymbol{r}} |^2 +\frac{\Tilde{g}}{2}|\psi_{\boldsymbol{r}} |^4),
\end{align}
thus it is of $\phi^4$ form.  $\mu$ is the chemical potential, and $\Tilde{g}=g/d^3$ is the effective interaction strength, rescaled by the discretization length  $d$.\\
The kinetic energy term accounts for coherent tunneling of Cooper pairs between neighboring sites and incorporates the Peirls coupling to the gauge field. It is given by:
\begin{align}
    F_{kin}
    =\sum_{\boldsymbol{r}} t \sum_j |\psi_{\boldsymbol{r}'(j)}-\psi_{r}\exp (i a_{j,\boldsymbol{r}})|^2 
\end{align}
where  $t$ is the tunnel coefficient and 
$a_{j,\boldsymbol{r}}$
the unitless vector potential, mentioned above.\\
The discretized form of the electromagnetic part is
\begin{align}
    F_{em}
    =\sum_{\boldsymbol{r}} \frac{\hbar^2}{4\mu_0 e^2 d} \sum_j 1-\cos ( \frac{-2ed}{\hbar} B_{j,\boldsymbol{r}} ),
\end{align}
where $\mu_0$ is the vacuum permeability, and $B_{i,\boldsymbol{r}}$  the local magnetic field components defined on the plaquettes of the lattice.
The magnetic field $B_{i,\boldsymbol{r}}$ is computed via 
$B_{i,\boldsymbol{r}} = \epsilon_{ijk} \delta_j A_{k,\boldsymbol{r}}$,
where $\delta_j$ is the discrete lattice derivative $\delta_j A_{k,\boldsymbol{r}}=(A_{k,\boldsymbol{r}'(j)}-A_{k,\boldsymbol{r}})/d_{j,\boldsymbol{r}}$ in the $j$-th direction.\\

As a key quantity, we consider the single-particle correlation function 
\begin{equation}
    \mathcal{C}(s)=\expval{\psi(r)^\dag W_{\boldsymbol{r},\boldsymbol{r+s}}\psi(r+s)}
    \label{eq:corr_func}
\end{equation}
of charged bosons, which includes the Wilson line or gauge-string $W_{\boldsymbol{r},\boldsymbol{r+s}}=\exp{i\sum_{l=0}^{s} a_{\boldsymbol{r+l}}}$. This non-local quantity must be included to ensure gauge invariance. Therefore the values of the vector potential along the path between the two points are accumulated like shown in \autoref{fig1} for $W_{\boldsymbol{r_1},\boldsymbol{r_2}}$  in purple. 
At large separations $s$, the correlation function asymptotically approaches the condensate density $n_0$. We identify the condensed and non-condensed state via the asymptotic behavior of this correlation function. If it displays long-range order, the condensate density $n_0$ is non-zero.\\

\subsection{UPDATE STRATEGY}
In the Monte Carlo simulation presented here, we perform three types of update steps. As we show below, this update strategy results in an acceptable autocorrelation. We note that the numerical effort of this Monte Carlo simulation is significantly larger than that of a U(1) theory of neutral bosons, due to the gauge string included in the defining correlation function.\\
The first update is a Metropolis updates of individual field values. Specifically, for the amplitude and phase of the order parameter a lattice site is randomly selected, and a trial update step $\Delta\phi$ for the phase or $\Delta\abs{\psi}$ for the amplitude of the order parameter is generated. The step is accepted if the resulting change in free energy $\Delta F$ satisfies the Metropolis criterion.\\
For the gauge field Metropolis updates, pairs of values $a_{j,\boldsymbol{r_1}} + \Delta a$, $a_{j,\boldsymbol{r_2}} - \Delta a$ (with $j \in {x, y, z}$) are simultaneously updated at positions $\boldsymbol{r_1},\boldsymbol{r_2}$ sharing the same coordinate in the ($k\neq j$)-th directions.
Via this pairwise update the sum over the closed loop of gauge fields in each direction satisfies the condition $\sum_{i=1}^{n_{\text{latt}}} a_j(j=i) = 2\pi n$ thus preserving the gauge invariance and the periodic boundary conditions. \\

The second update step is a layer update  applied to the gauge fields. In each attempted update step, all $a_j$ values at a fixed position $j = m$ were increased by $\Delta a$, while those at $j = n$ were decreased by the same amount. A figure to illustrate the layer updates is shown in the Supplemental material \cite{supMat}. As a result, a total of $2n_{\text{latt}}^2$ gauge field components are updated simultaneously, enabling large  steps of the system. An additional advantage of this strategy is that the magnetic fields, and consequently the electromagnetic free energy $F_{\text{em}}$, remain unchanged during the update. This increases the likelihood of accepting this update for large steps $\Delta a$, because the large electromagnetic free energy $F_{\text{em}}$ suppresses  individual fluctuations of the $a$-field components.\\

The third update is a zero-energy update step. In these updates, individual values of the amplitude and the phase of the order parameter, as well as pairs of gauge field components, similar to the pairwise Metropolis update described before, are exchanged in a way that kept the total free energy, see Supplemental material \cite{supMat}.\\

This combination of update strategies enables us to reduce the large computational effort involved in including the gauge field to $10^7$ Monte-Carlo update steps per simulated grid point. The details of the numerical simulation procedures as well as the analysis of the autocorrelation can be found in the Supplemental material \cite{supMat}.\\

\subsection{RESULTS}
In the following, we present our numerical results obtained from the Monte-Carlo simulation of our lattice gauge model. 
In \autoref{tab:parameter} we show the two parameter sets that we work with. Given the large numbers of parameters of the model, we focus on these two examples here. We note that the large magnitude of the prefactor of $F_{em}$, derives from the approximate size of the unit cell of the simulation, see also \cite{Homann2021}. The two parameter sets differ in terms of the magnitude of $g$ and $\mu$, while the ratio is kept fixed. Therefore the average density is approximately the same for both sets, because $|\psi_0|^2 = \mu/g$. However, the key difference between the parameter sets is the magnitude of the density fluctuations. Because $g$ is four orders of magnitude larger in parameter set two, the density fluctuations are noticeably suppressed. We note that these parameters give rise to a type-II superconductor.
If not explicitly stated the results are from parameter set one.
All temperatures will be given in unitless form in dependence of the tunnel coefficient as $\Tilde{T}=k_B T /t_h$.\\

\paragraph{Vortices.-}
We will first look into the microscopic vortex fluctuations. The vorticity in our discretized  system with gauge fields is defined on the plaquettes as the phase winding of a closed path
\begin{align}
    \nu=\oint \left( \nabla\phi +\frac{2e}{\hbar}\bold{A}\right)dr -\oint \frac{2e}{\hbar}\bold{A}dr
\end{align}
\cite{Homann2024}. For each plaquette the vorticity is an integer number, where positive numbers correspond to vortices and negative numbers to antivortices.
Snapshots of the 3D system with vortices (blue) and antivortices (red) are shown in \autoref{fig2} for small systems with a grid  of  $4^3$  points at increasing temperatures, and in the Supplemental material  \cite{supMat} for a grid of  $14^3$  points.
For smaller temperatures (\autoref{fig2}~(a)) isolated vortex rings emerge. For increasing temperature the number of vortices grow and more complex vortex clusters form (\autoref{fig2}~(b) and (c)). Further double vortices with $\nu=\pm 2$ emerge. 
While the number of vortices and double vortices increases with the temperature,  the rate of increase reaches its maximum at the critical temperature. The increase of the average vortex density and the double vortices is depicted together with the spatial vortex correlation function in the Supplemental material \cite{supMat}.\\ 

\begin{figure}
    \centering
    \includegraphics[width=1\linewidth]{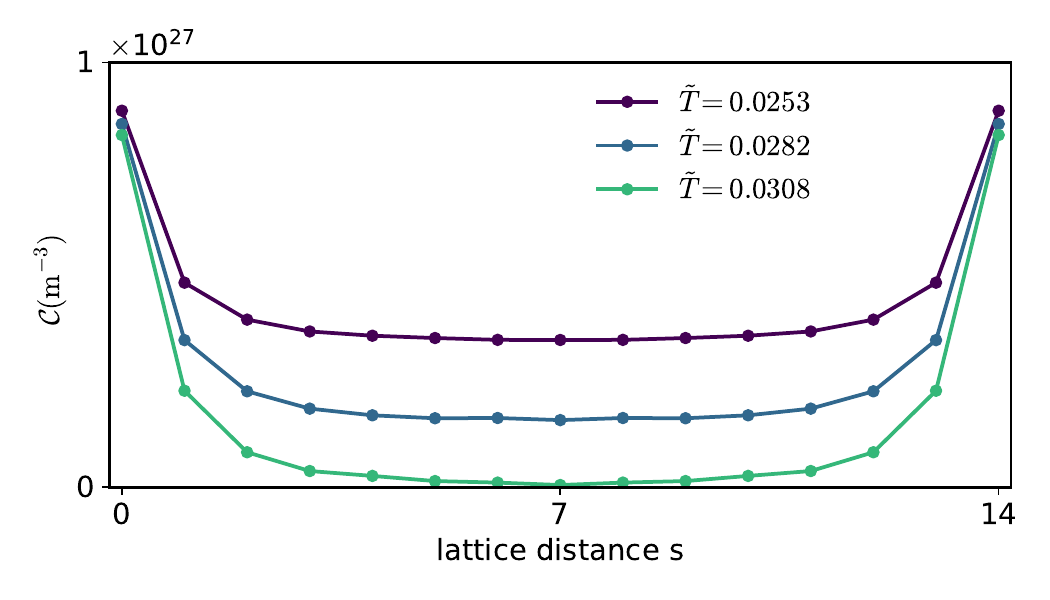}
    \caption{
    Single particle correlation function $\mathcal{C}(s)$ with Wilson line as a function of the distant in grid points $s$. For $T<T_c$ the correlation function converges to a nonzero Cooper pair density $n_0$ ($\Tilde{T}=0.0253$ and $\Tilde{T}=0.0282$ in the plot). For $T>T_c$ ($\Tilde{T}=0.0308$ in plot) only short-range correlations are present and $n_0\rightarrow0$. The recovery of the correlation function for large distances $s$ reflects the periodic boundary conditions combined with the gauge invariance.
    }
    \label{fig3}
\end{figure}

\paragraph{Cooper pair density.-}
To characterize the spatial coherence and the Cooper pair density we will next study the gauge-invariant single-particle correlation function $\mathcal{C}(s)$. \autoref{fig3} shows the behavior of the single-particle correlation function $\mathcal{C}(s)$ (\autoref{eq:corr_func}) as a function of distance $s$ in the x-direction of the system in units of lattice sites for three different temperatures. Above the critical temperature, at  $\Tilde{T}=0.0308$ in the plot, the system exhibits only short-range order, and the Cooper pair density $n_0$ vanishes. In contrast, at $\Tilde{T}=0.0253$ and $\Tilde{T}=0.0282$  below the critical temperature, long-range order is clearly observed and the correlation function approaching  asymptotically the value $n_0$.
Due to the periodic boundary conditions of the finite lattice system, the correlation function shows a recurrence at large distances.
This behavior reflects the gauge invariance under periodic boundary conditions of the  system.\\

To determine the Cooper pair density $n_0$ more precisely, finite-size scaling \cite{Malthe-Sørenssen2024} of the sum of the single-particle correlation function between all pairs of lattice sites are evaluated for various cubic lattice sizes $n_{\text{latt}}^3$ with $n_{\text{latt}}\in[4, 14]$. 
The contribution from short-range correlations scales as $(r_0/n_\text{latt})$ and the long range correlation corresponds to the Cooper pair density where $r_0$ is the correlation length of $C(s)$.
By fitting the averaged correlation sum to 
$(1/n_{\text{latt}}^6) \sum_{i,j} \langle \psi^\dagger(\boldsymbol{r}_i) W \psi(\boldsymbol{r}_j) \rangle \approx \left( r_0/n_{\text{latt}} \right) + n_0$, 
the Cooper pair density $n_0$ can be extracted.\\

\begin{figure}
    \centering
    \begin{overpic}[width=\linewidth]{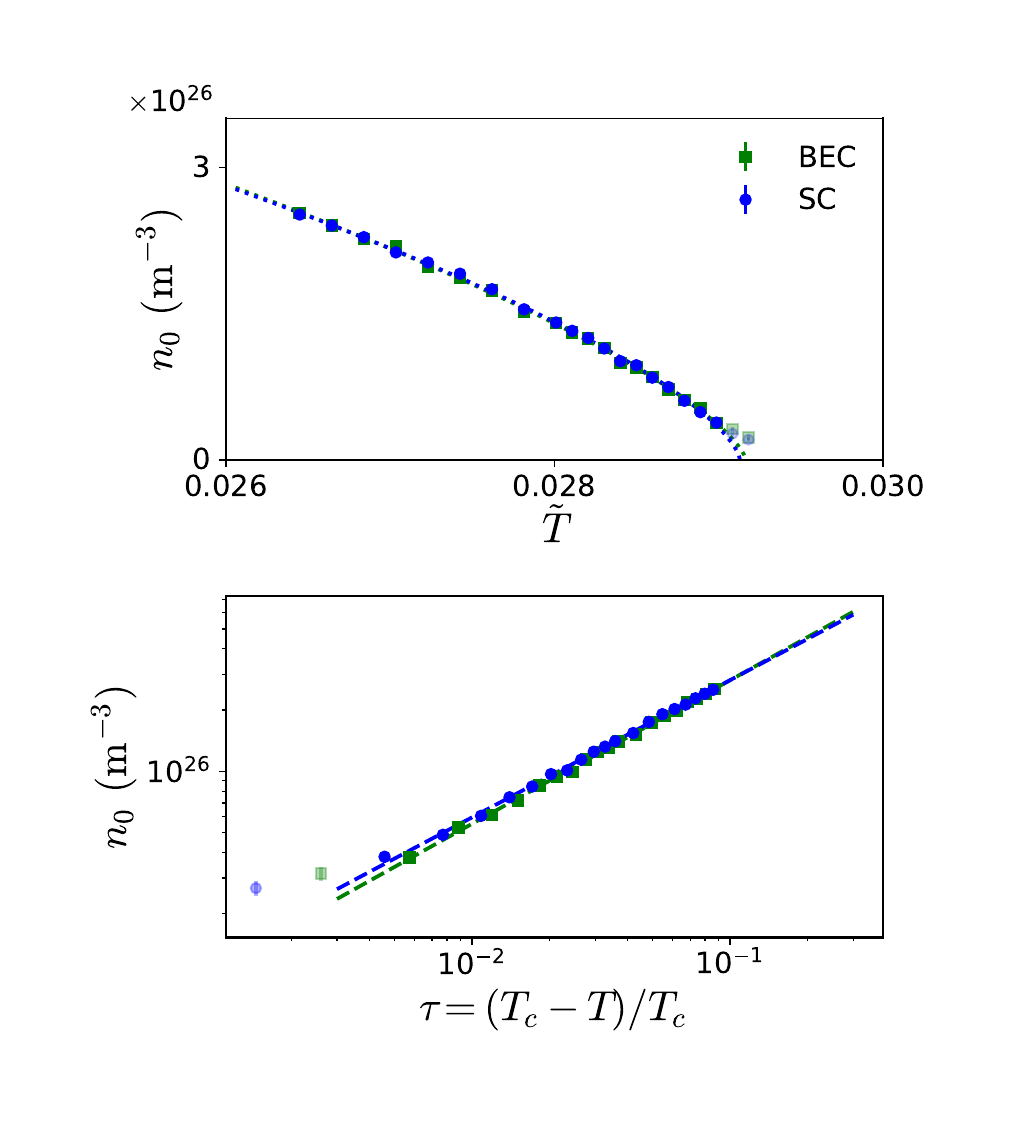}
    \put(0,93){(a)}
    \put(0,43){(b)}
    \end{overpic}
    \caption{
    (a) The Cooper pair density $n_0$ as a function of the unitless temperature $\Tilde{T}$ for the superconducting case, i.e. in  the U(1) gauge theory (SC, blue)  and  the condensate density of a Bose-Einstein condensate without gauge field (BEC, green) for comparison for one of the evaluated temperature intervals. 
    Near the phase transition finite size effects causes a smearing of the phase transition, these transparent points are not used for the fit. Through the non-transparent data points a fit of the order parameter and the critical temperature is shown (dashed lines).  \\
    (b) Logarithmic plot of the Cooper pair density $n_0$ as a function of reduced temperature $\tau=(T_c-T)/T_c$ for the same fitting interval as above. The critical exponent extracted from all evaluated temperature intervals for the superconducting case with gauge string, $\beta_{\text{U(1) gauge}}=0.344 \pm 0.014$, is consistent with the BEC case and the expected value for the  U(1) universality class. 
    }
    \label{fig4}
\end{figure}
In \autoref{fig4}~(a), the Cooper pair density is plotted in blue as a function of temperature. For comparison, the condensate density of a Bose-Einstein condensate (BEC) without a gauge field for the same parameter set is shown in green. This is obtained by setting $A_j(\boldsymbol{r})$ to zero, which gives a U(1) theory instead of the U(1) gauge theory studied throughout this paper.
Near the phase transition finite size effects causes a smearing out of the condensate density, compared to the thermodynamic limit, for the BEC and the U(1) gauge case. These points are plotted as transparent and are not used for the following fits.
We fit the critical exponent $\beta$ and the critical temperature $T_c$ using the relation $n_0 \propto ((T-T_c)/T_c)^{2\beta}~$ (dashed lines) for several temperature fitting intervals and obtain $\beta_{\text{U(1) gauge}}=0.344 \pm 0.014$. 
We note that this fitting approach displays a dependence on the fitting interval that is used, as discussed in the Supplemental material \cite{supMat}. The resulting systematic error is of the order of $\Delta \beta \approx 0.014$ for the U(1) gauge critical exponent. 
The fit of the largest temperature interval is shown in \autoref{fig4}.
For comparison, the case of a BEC of neutral bosons yields $\beta_{\text{BEC}} = 0.351\pm 0.011$. 
The average obtained critical temperature of the superconductor $\Tilde{T}_{c,\text{U(1) gauge}}=0.02891 \pm 0.00003$ does not differ significantly from the average critical temperature $\Tilde{T}_{c,\text{BEC}}=0.02894 \pm 0.00003$ of the BEC. \\

In \autoref{fig4}~(b), the condensate density and the corresponding fit are shown for the same temperature interval as a function of the reduced temperature $\tau = (T_c - T)/T_c$, demonstrating the agreement between the fitted critical exponent and critical temperature and the numerical data. 
The plot of the determination of the critical exponent and the critical temperature  with the condensate density for the second parameter set with smaller density fluctuations is shown in the Supplemental material \cite{supMat} and delivers consistent values for $\beta$.
Additionally we computed $\beta$ via the minimum of the least square error of fits for fixed $\beta$. These values are also given in the Supplemental material \cite{supMat} and are in agreement with the values obtained from the direct fits of $\beta$.
The obtained critical exponents  $\beta_{\text{U(1) gauge}}$ matches with the critical exponent associated with the U(1) universality class in three dimensions \cite{Guida1998} which we have reproduced here with $\beta_{BEC}$.\\

\paragraph{Anomalous dimension $\eta$.- }
The magnitude of the anomalous dimension $\eta$ in U(1) gauge systems has been discussed controversially in Refs. \cite{Hove2000,Prokofev2006,Hove2006}.
In our simulation, we determine the anomalous dimension through the finite size scaling of the correlation function.  Instead of fitting $\mathcal{C}(x)=\expval{\psi_i^\dag W \psi_{i+r}}$ with the fitting function $Aexp(-r/\xi(\tau))/r^{1+\eta}+n_0$, we calculate the least square error of the best fit for fixed $\eta$ using the cumulated correlation function. For this purpose, we take the lattice size $n_l^3=18^3$ and evaluate the sum of all correlation functions $\mathcal{C}(r_i,r_j)$ with $|r_i-r_j|<|r|_{max}$ for $r_{\text{max}}=[3,9]$ and fit it to
\begin{align}
    &\frac{1}{n_l^3} \frac{3}{4\pi |r|_{max}^3}\sum_i \sum_{j, |r_i-r_j|<|r|_{max}}  \mathcal{C}(r_i-r_j)\\
    &=\frac{3A}{|r|_{max}^3} \xi(\tau)^{2-\eta}\Gamma(2-\eta,\frac{|r|_{max}}{\xi}).
\end{align}
\begin{figure}
    \centering
    \includegraphics[width=1\linewidth]{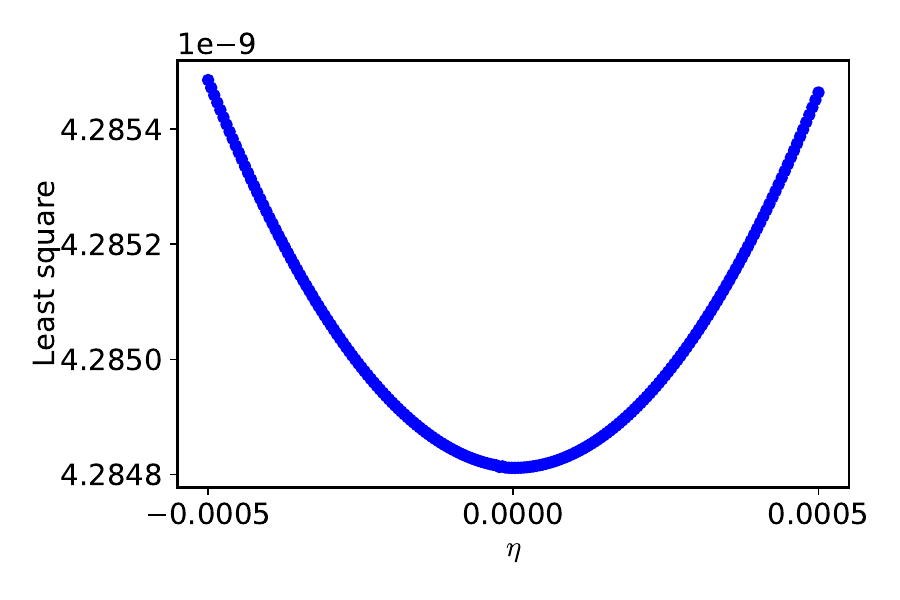}
    \caption{The least square error of the fits to the sum of the correlation functions $\sum \mathcal{C}(|r_i-r_j|<r_{\text{max}})$ as a function of $r_{\text{max}}$ for $\Tilde{T}= 0.0282$  and parameter set 1. The exponent $\eta$ is kept fixed for every single fit. The $\eta$ with the best fit and therefore smallest least square error is in the range $\eta \approx (-1,1)\times10^{-4}$ for the different temperatures and parameter sets.
    }
    \label{fig_eta}
\end{figure}
In \autoref{fig_eta} we show a typical plot for the least square error of the fit as a function of $\eta$ for $\Tilde{T}=0.0282$. The fits for additional temperatures and the second parameter set are shown in the Supplemental material \cite{supMat}. The minimum of the fitting error for our parameter sets indicate an anomalous dimension $\eta$ in the interval $ (-1,1)\times10^{-4}$ , which is significantly smaller than the $\eta_{\text{U(1)}}=0.038$ \cite{Campostrini2001}  obtained for the U(1) universality class.
\\

\paragraph{Heat capacity.-}
We now focus on the critical signature of the heat capacity.
Duality arguments indicate that the heat capacity of type-II-superconductor in the limit of constant density and strong gauge field coupling creates an inverted XY-transition \cite{Dasgupta1981}. 
We therefore evaluate the heat capacity for the two parameter sets of our lattice gauge model.  The first parameter set of \autoref{tab:parameter} is not suitable for a Villain approximation as the fluctuation of the density per site is temperature dependent up to 37\%. In contrast, the second parameter set has only 0.33\%  fluctuations at maximum allowing approximately a Villain approximation. As mentioned above, the second parameter set differs from the first parameter set only by a factor of $10^4$ for the chemical potential $\mu$ and the interaction strength $g$, thus suppressing density fluctuations. \\

\begin{figure}
    \centering
    \begin{overpic}[width=\linewidth]{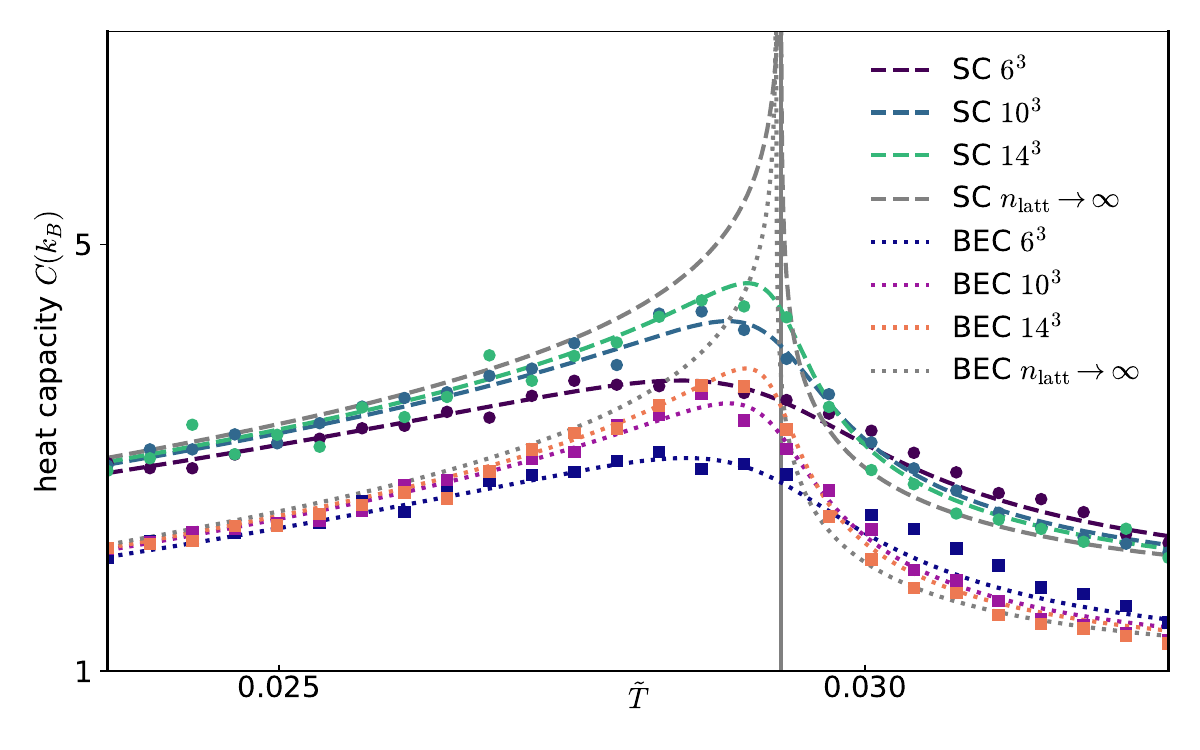}
    \put(0,53){(a)} 
    \end{overpic}
    \begin{overpic}[width=\linewidth]{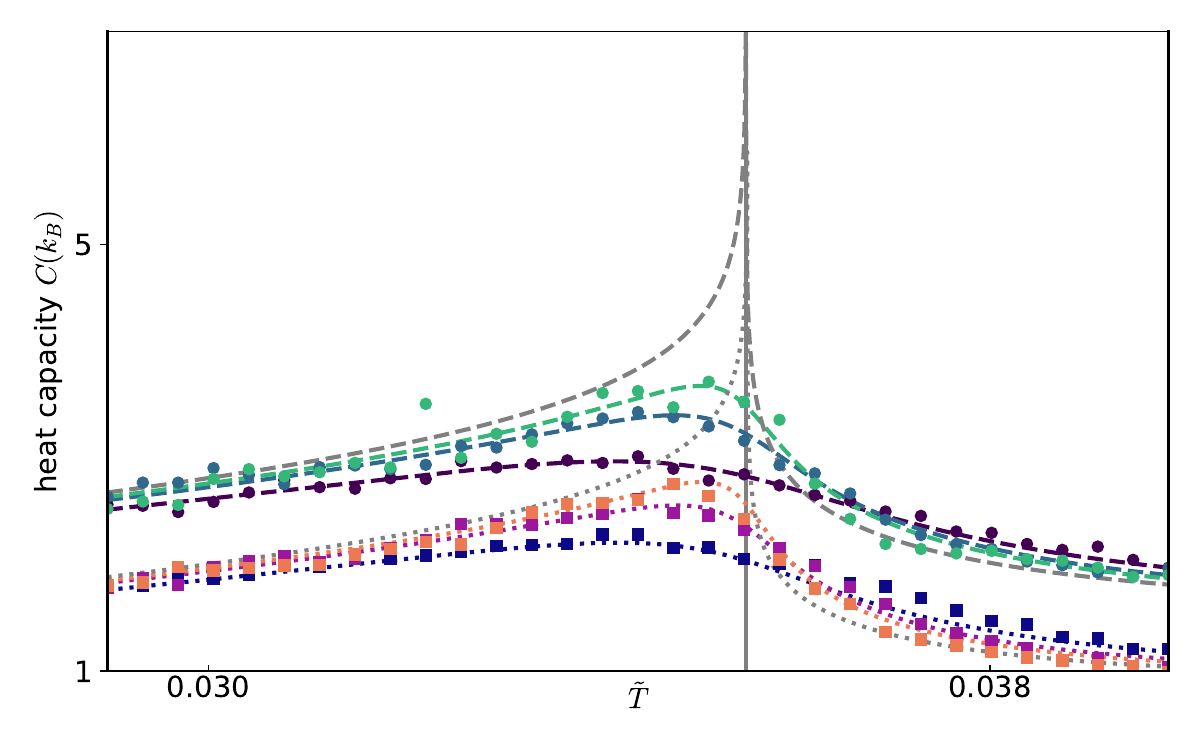}
    \put(0,53){(b)}
    \end{overpic}
    \caption{ 
    Heat capacity $C$ in units of $k_B$ as a function of the unitless temperature $\Tilde{T}$ for different lattice sizes with fits for the superconducting case with gauge field (SC, data=circles, fit=dotted lines) and the gauge-free BEC case (BEC, data=squares, fit=dashed lines).
    In addition plotted in grey is the extrapolation to infinite lattice sizes. Shown in (a) is the heat capacity for the parameter set 1 of \autoref{tab:parameter} with large fluctuations. In (b) the heat capacity of the parameter set 2 with only small fluctuations in the particle density per site is shown, proving that the regime of small density fluctuations does not change the behavior. 
    The heat capacity of both parameter sets have a clear XY-transition.
    }
    \label{fig5}
\end{figure}

\begin{table}[h!bt]
    \begin{tabular}{p{5.5cm} p{1.3cm} p{1.3cm}}
        parameter set & 1 & 2\\
        \hline
        \hline
         $\mu$ (eV) & 10.0 & $10^5$\\ 
         $\Tilde{g}$ (eV)& $1250$& $1.25\cdot 10^7$ \\
         $g$ (eV $\text{\AA{}}^3$ ) & $10^4$ & $10^8$\\
         $t_h$ (meV) & 238.0 & 238.0\\
         $d$ (\AA{})& 2.0& 2.0\\
         $\hbar^2/(4\mu_0 e^2 d)$ (eV) & $2690$& $2690$\\
         \hline
         \hline
    \end{tabular}
    \caption{Model parameter sets used for the simulation. 
    }
    \label{tab:parameter}
\end{table}

The heat capacity is obtained from the energy fluctuations according to $C=1/k_B T^2 (\expval{E^2}-\expval{E}^2)$ \cite{Pathria1996}.
For the first parameter set the heat capacity is depicted in \autoref{fig5}~(a) for the lattice sizes $n_{\text{latt}}^3$ of $6^3$, $10^3$ and $14^3$ grid points (SC). For comparison the heat capacity of the same parameter set without gauge field (BEC) is plotted in the same plot.
The heat capacities of the second parameter set with small density fluctuations is shown in \autoref{fig5}~(b).
We fitted the heat capacity $C$ of all three lattice sizes to the finite-size scaling function
\begin{align}
    C(T)=& -\frac{A}{2} \log ((\tau+a n_{\text{latt}}^{-3/2})^2+b^2n_{\text{latt}}^{-3})\\
    & -\frac{DA}{\pi}\arctan(\frac{\tau n_{\text{latt}}^{3/2}+a}{b})+B_0+B_1(T-T_c) ,
\end{align}
where the small critical exponent $\alpha$ of the heat capacity is set to zero \cite{Dasgupta1981}. The fits are plotted in \autoref{fig5} for the superconductor with gauge fields (SC, dashed line) and for the condensate of neutral bosons (BEC, dotted lines) as well as the extrapolation for infinite lattice sizes in grey.
The resulting critical temperatures $\Tilde{T}_c$ are in the standard deviation of the critical temperatures obtained from the condensate density. \\

The figures show clearly, that both parameter sets have a typical XY-transition.
Furthermore, the heat capacity for the superconductor model with gauge field is larger than for the BEC case without gauge fields due to the additional degrees of freedom of the gauge field. Moreover, the heat capacity for the parameter set one with large fluctuations in \autoref{fig5}~(a) is larger than for the parameter set with neglectable fluctuations in \autoref{fig5}~(b), corresponding to the degree of freedom of the amplitude of the order parameter.\\

\subsection{CONCLUSION}
In this work, we have determined the critical behavior of a U(1) lattice gauge system, as a model for unconventional superconductors. Given the gauge nature of the model, we include a gauge string in the definition of the order parameter correlation, making it a non-local, multi-site correlation function, in contrast to the two-point correlation function of a U(1) system of neutral bosons. We map out the correlation function for the lattice gauge model, and determine the critical exponent $\beta$. Remarkably, despite the conceptual difference of the gauge model and the non-gauge model, and the difference of the nature of the correlation function, the exponent $\beta$ of the gauge model is consistent with the exponent of the non-gauge model, i.e. of condensation of neutral bosons. 
Further, we mapped out the anomalous dimension $\eta$ and found it to be in the range of $(-1,1)\times10^{-4}$ for our parameter sets, which is smaller than the value of $\eta$ for the condensation transition of a U(1) system without a gauge field. As another critical signature, we generate the temperature dependence of the heat capacity across the phase transition, which is consistent with an XY transition. Furthermore, we determine the vortex activation across the transition, where we observe the maximal increase near the critical temperature.

With these signatures of the phase transition, we have identified the defining properties of the critical behavior of the U(1) lattice gauge system in three dimensions, which will impact the modeling and understanding of unconventional superconductors.

\subsection{Acknowledgment}
We thank Flavio Nogueira for inspiring discussions.\\
We acknowledge support from the Cluster of Excellence 'CUI: Advanced Imaging of Matter' of the Deutsche Forschungsgemeinschaft (DFG) - EXC 2056 - project ID 390715994 and ERDF of the European Union and by ’Fonds of the Hamburg Ministry of Science, Research, Equalities and Districts (BWFGB)’.

\subsection*{Data Availability}
The data that support the findings of this article are openly available \cite{dataset}.

\FloatBarrier

%

\clearpage
\newpage

\onecolumngrid
\appendix
\renewcommand{\thesection}{\Roman{section}}
\renewcommand{\thesubsection}{\Roman{subsection}}

 \begin{center}
{\Large\textbf{Supplemental Material:
Critical behavior of the thermal phase transition of U(1) lattice gauge systems}}
\end{center}

First, we present in \autoref{subsec:MC_parameter} the details of our Monte Carlo update strategy,  all simulation parameters and the autocorrelation plots. 
Second, we show in \autoref{subsec:beta_set2} the fitting values of $\beta$ for all evaluated temperature intervals and the fit of the critical exponent $\beta$ for the second parameter set to confirm the agreement to the U(1) value.
Further, we show the evaluation of $\beta$ via the least square error for the same parameter sets and temperature intervals.
Furthermore, in \autoref{subsec:vortices} we present the plot of the vortex density over temperature, the spatial vortex correlation function and a screenshot of the vorticity on the $14^3$ lattice. 
Last, we show in \autoref{subsec:ana_dim} the least square error as a function of the anomalous dimension $\eta$ for a second temperature of parameter set 1 and two temperatures of the  parameter set 2.\\

\subsection{Monte Carlo update strategy and Simulation parameter}\label{subsec:MC_parameter}

For the simulation with gauge field, the ratio of update strategies was $76.95\%$ Metropolis update steps, $13.05\%$ layer updates, and $10\%$ zero-energy change updates. The standard deviations of the Gaussian distributions used for the proposed Metropolis updates were set for the first parameter set to
$\sigma_{\abs{\psi},1}=1/300+T/600$ and $\sigma_{\phi}=\pi /10$ for the amplitude and phase of the order parameter $\psi$, and  $\sigma_a=14/3*10^{-4} +T*10^{-4}/3 $  for the components of the vector potential $a_i$, to ensure an acceptance rate of approximately $50\%$ for all temperatures $T$. For the column-wise updates of the vector potential, the standard deviation was set to $\sigma_{a,\text{layer}}= \pi /40$ .
The standard deviation for the second parameter set differs only in $\sigma_{\abs{\psi}, 2}=(1/300+T/600)/81$. \\

\begin{figure}[h]
    \centering
    \includegraphics[width=0.5\linewidth]{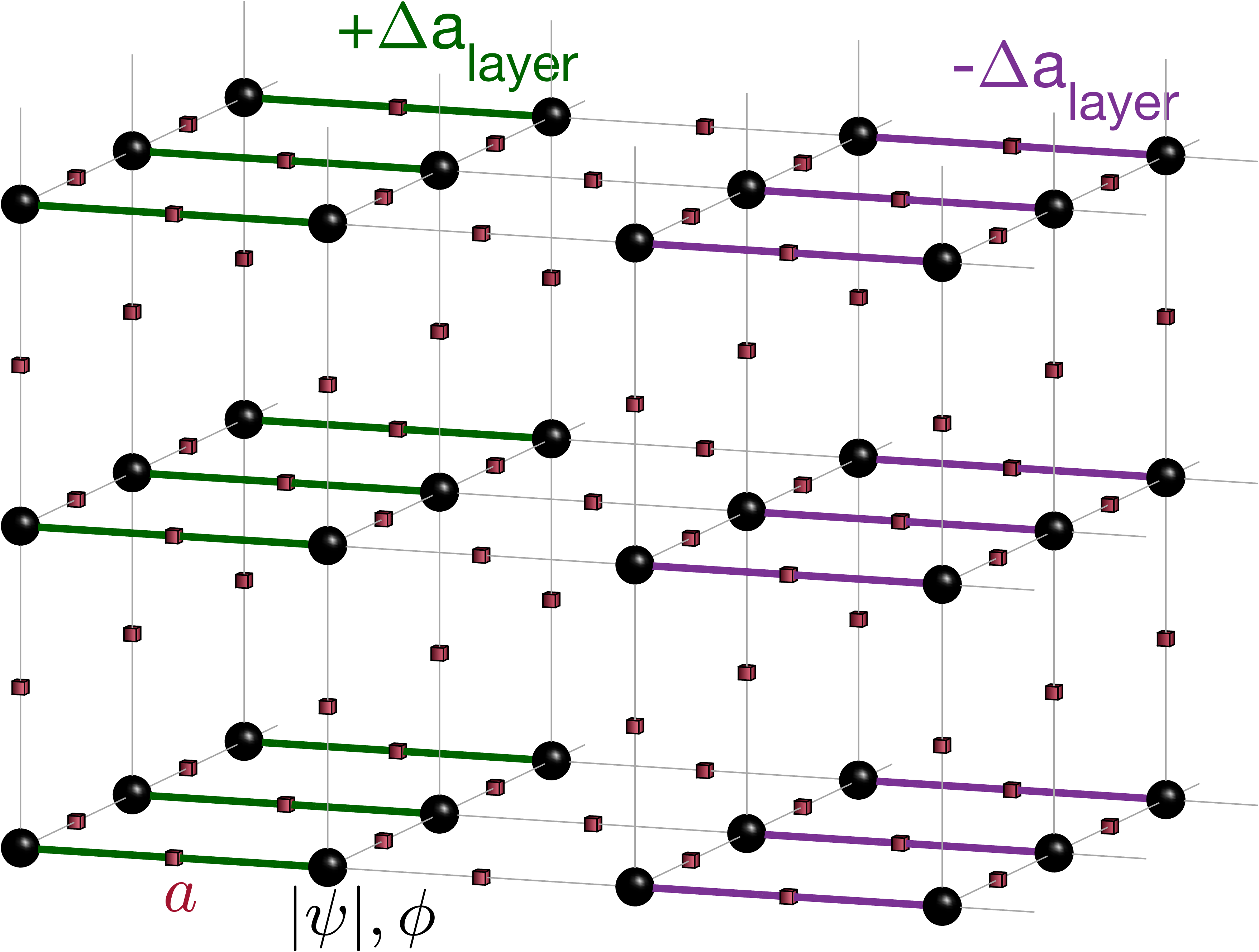}
    \caption{For the layer updates of the gauge fields all $a_j$ values at  position $j = m$ were increased by $\Delta a_{\text{layer}}$, while those at $j = n$ were decreased by $\Delta a_{\text{layer}}$.}
    \label{figSM1}
\end{figure}

A sketch explaining the layer update strategy of the gauge fields is depicted in \autoref{figSM1}. 
In each proposed update step, all $a_j$ values at a fixed position $j = m$ were increased by $\Delta a_{\text{layer}}$, while those at $j = n$ were decreased by the same amount.  As a result, a total of $2n_{\text{latt}}^2$ gauge field components were updated simultaneously. 

\begin{figure}
    \centering
    \begin{overpic}[width=0.45\linewidth]{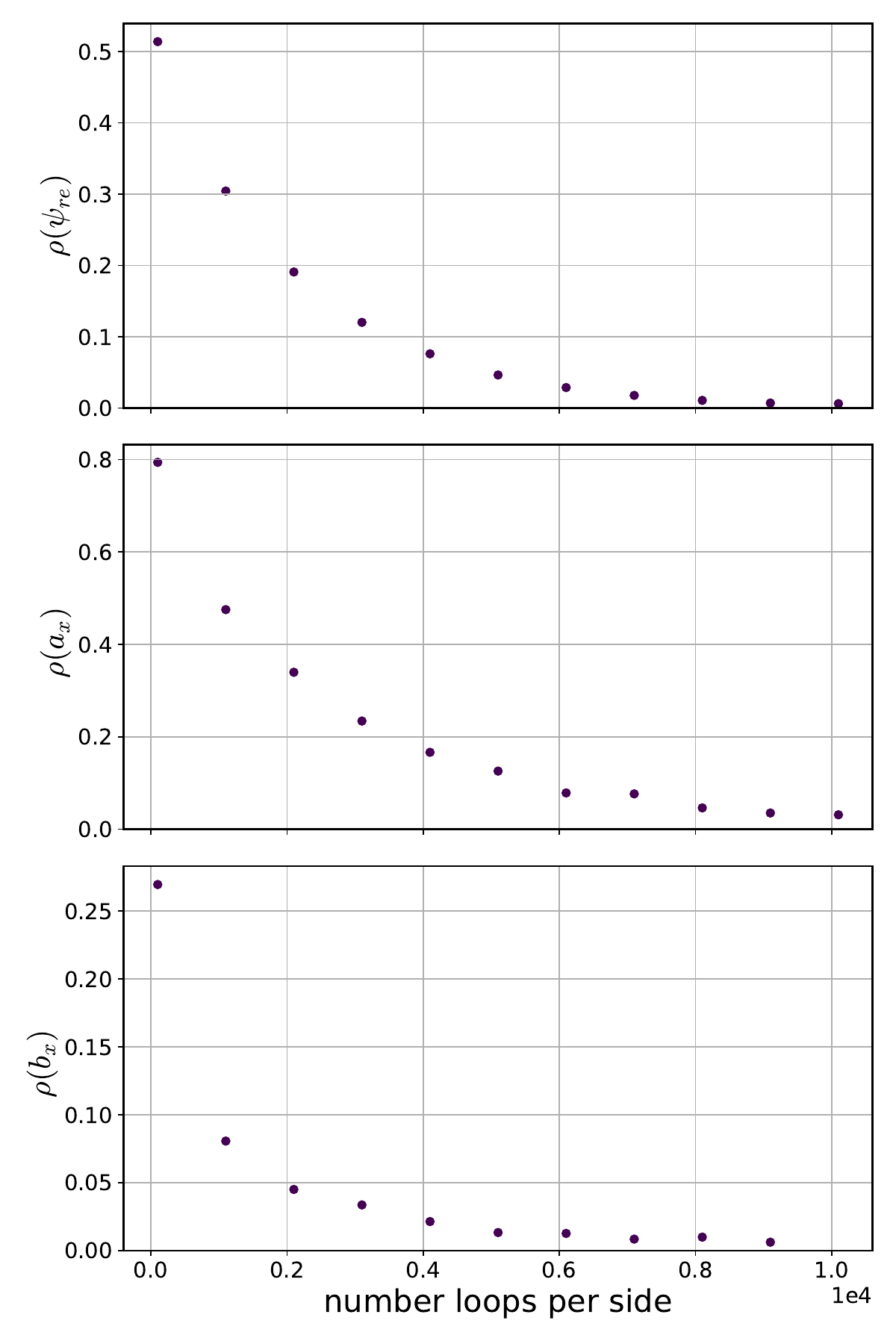}
    \put(0,97){(a)} 
    \end{overpic}
    \begin{overpic}[width=0.45\linewidth]{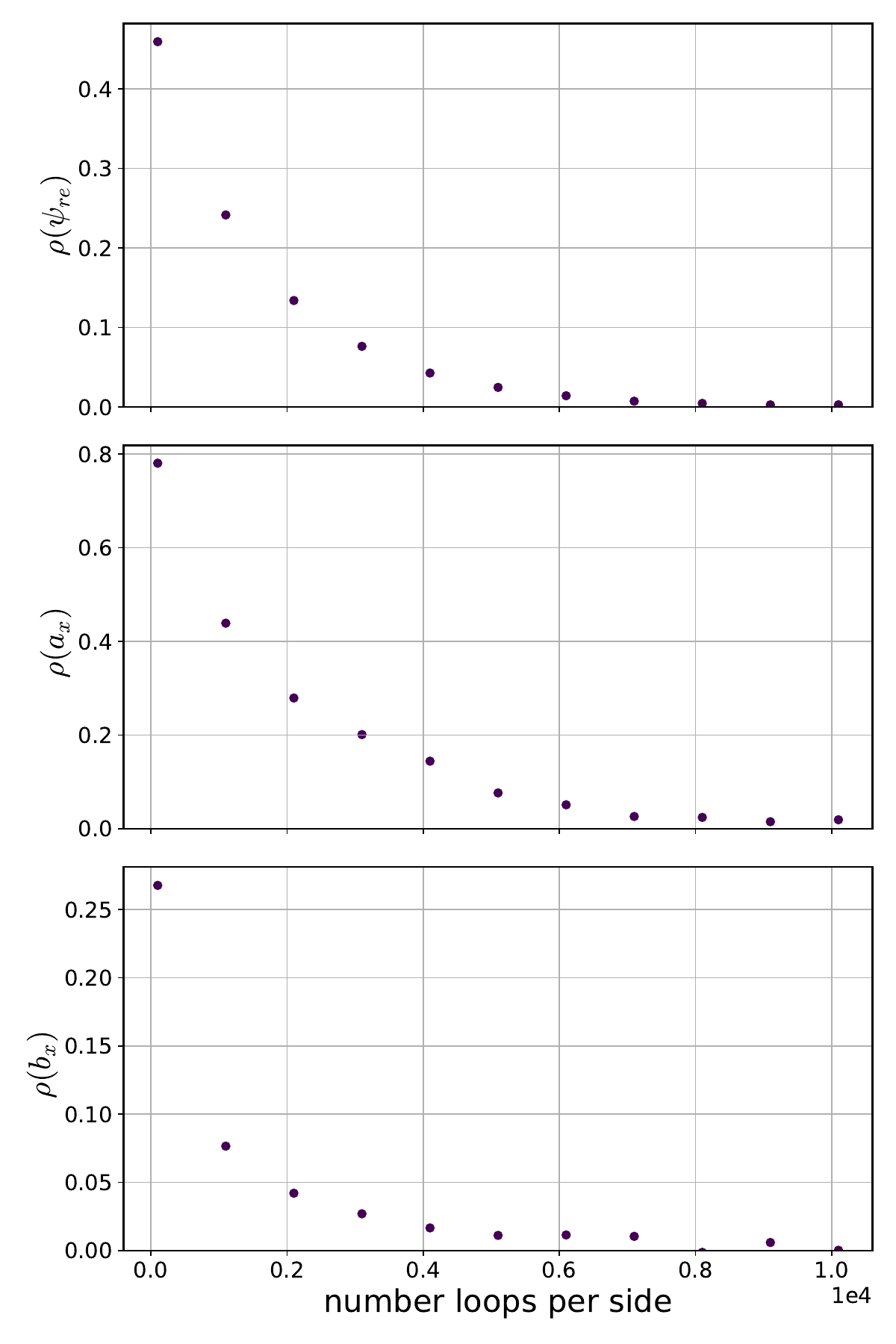}
    \put(0,97){(b)} 
    \end{overpic}
    \caption{ 
    The normalized autocovariance $\rho$ of the real valued order parameter $\psi_{re}$, the vector potential $a_x$ and the derived B-field values $b_x$ for (a) $\Tilde{T}=0.0235$ and parameter set one and (b) $\Tilde{T}=0.0290$ and parameter set two.  
    In the simulations in the main text  $10^7$ inner loops per side were performed between the readouts to ensure autocovariance of much smaller 1\%.
    }
    \label{figSM2}
\end{figure}

For the zero-energy updates, one of the fields $(\abs{\psi}, \phi, a_x, a_y, a_z)$ is randomly selected to be updated. In addition, for the fields ${\abs{\psi}, \phi}$ of the order parameter, a single site $\boldsymbol{r}=(x,y,z)$ is randomly selected. Then an optimization is performed to find a new value $\abs{\psi (\boldsymbol{r})}$ (or $\phi(\boldsymbol{r})$), which results in exactly the same free energy as the previous configuration.\\

For the zero-energy updates of the $a$-fields, gauge invariance under periodic boundary conditions needs to be conserved. If we select $a_j$ with $j \in (x,y,z)$ to be updated we therefore choose two sites $\boldsymbol{r}_1,\boldsymbol{r}_2$ to be updated to $a_{\boldsymbol{r}_1}+\Delta a$ and $a_{\boldsymbol{r}_2}-\Delta a$ which are on the same loop of the periodic boundary. So the coordinates $\boldsymbol{r}_1$ and $\boldsymbol{r}_2$ share the coordinates for $j \neq i$ and have different coordinate values in the $j=i$ direction. The four coordinate values defining $\boldsymbol{r}_1$ and $\boldsymbol{r}_2$ are chosen randomly. Then an optimization for a value $\Delta a$ that result in no change in the total free energy is performed and $\Delta a$ is applied to the two sites for the next Monte Carlo configuration. \\

The number of update steps per lattice side between measurements was $10^7$. In \autoref{figSM2}, the normalized autocovariance of the order parameter, the vector potential, and the magnetic field is shown as a function of the number of update steps at temperatures $\Tilde{T}=0.0235$ for parameter set one and $\Tilde{T}=0.0290$ for parameter set two, showing the achievement of an autocovariance much smaller than 1\% even for the smallest simulated temperatures.\\

For the comparison with the Bose-Einstein condensate without a gauge field, already $10^4$ update steps between measurements were sufficient to achieve negligible autocorrelation. The order parameter field was updated using $90\%$ Metropolis steps and $10\%$ zero-energy updates. \\

For the plot of the condensate density over the temperature of parameter set one in the main text 400  Monte Carlo measurements were taken per temperature and lattice size. For the plot of parameter set two in the appendix 200  Monte Carlo measurements were taken each. The evaluated temperature ranges are specified in \autoref{subsec:beta_set2}.
For parameter set one the data points were spaced by  0.00018 below $\Tilde{T}=0.02788$ and spaced by 0.00009 above $\Tilde{T}=0.02788$.
For the BEC data the same number of data points and measurements were used. 
For the parameter set two  an unitless temperature spacing of 0.00036 was used.\\

For the plot of the heat capacity of the parameter set one the unitless temperatures [0.0235,0.0326] and for the parameter set two [0.0290,0.0398] were taken with a spacing of 0.00036. For the lattice sizes $6^3$ and $10^3$ 5000 configurations were evaluated. For the lattice size $14^3$ 1000 configurations were considered per temperature. \\

\subsection{Systematic error $\Delta \beta$ and condensate density plot parameter set two} \label{subsec:beta_set2}

The fitting approach of the critical exponent $\beta$ and the critical temperature $T_c$ displays a dependence of the fitting interval. We therefore use several fitting intervals, to obtain our exponent $\beta$ and to get an estimate of the systematic error $\Delta\beta$. 
We listed all fitted exponents and critical temperatures of parameter set one in \autoref{tab:beta_power0}. The fitting errors of the individual temperature intervals are much smaller as the systematic error due to the chosen temperature interval, so we will neglect them here. 
The combined result of all temperature intervals for parameter set one are $\beta_{\text{U(1) gauge}}=0.344 \pm 0.14$,  $\Tilde{T}_{c,\text{U(1) gauge}}=0.02891 \pm 0.00003$, $\beta_{\text{BEC}}=0.351 \pm 0.11$ and $\Tilde{T}_{c,\text{BEC}}=0.02894 \pm 0.00003$. \\

\begin{table}
    \centering
    \begin{tabular}{p{2.0cm} p{2.0cm} p{2.0cm} p{2.0cm}p{2.0cm}p{2.0cm}}
        $T_{\text{start}}$ & $T_{\text{end}}$ & $\beta_{\text{U(1) gauge}}$ & $T_{c, \text{U(1) gauge}}$ & $\beta_{\text{U(1) }}$ & $T_{c, \text{U(1)}}$\\
        \hline \\
         0.02643 &0.02860 & 0.322 & 0.02887 & 0.340 & 0.02891\\
         0.02679&0.02860   & 0.329 &0.02888& 0.346& 0.02892\\
         0.02716& 0.02860  & 0.350 &0.02892&0.335&0.02890 \\
         0.02752& 0.02860  & 0.348 &0.02892 &0.354&0.02893 \\
         0.02643 &   0.02878  &0.337&0.02892 &0.352& 0.02895\\
         0.02679 &   0.02878  & 0.344&0.02893 &0.359&0.02896 \\
         0.02716&   0.02878  &0.361&0.02895&0.356&0.02896 \\
         0.02752 & 0.02878  & 0.364 &0.02896&0.372&0.02898 \\ 
    \end{tabular}
    \caption{The parameters $\beta$ and $T_c$ for the U(1) and the U(1) gauge lattice of the parameter set one fitted for eight different temperature intervals. The combined result of all temperature intervals for parameter set one are  $\beta_{\text{U(1) gauge}}=0.344 \pm 0.14$ , $\Tilde{T}_{c,\text{U(1) gauge}}=0.02891 \pm 0.00003$, $\beta_{\text{BEC}}=0.351 \pm 0.11$ and $\Tilde{T}_{c,\text{BEC}}=0.02894 \pm 0.00003$.}
    \label{tab:beta_power0}
\end{table}

In addition to the direct fit of $\beta$, we fit for each temperature interval $\beta$ via the minima of the least square error analogously to the fitting approach of $\eta$. The results for each temperature interval of parameter set 1 are listed in \autoref{tab:beta_power0_LS}, resulting in $\beta_{\text{U(1) gauge}}=0.345 \pm 0.013$,
$\Tilde{T}_{c,\text{U(1) gauge}}=0.02888 \pm 0.0001$,  $\beta_{\text{BEC}}=0.354 \pm 0.11$ and $\Tilde{T}_{c,\text{BEC}}= 0.02894\pm 0.00002$. These values are in agreement to the direct fit of $\beta$.\\

\begin{table}
    \centering
    \begin{tabular}{p{2.0cm} p{2.0cm} p{2.0cm} p{2.0cm}p{2.0cm}p{2.0cm}}
        $T_{\text{start}}$ & $T_{\text{end}}$ & $\beta_{\text{U(1) gauge}}$ & $T_{c, \text{U(1) gauge}}$ & $\beta_{\text{U(1) }}$ & $T_{c, \text{U(1)}}$\\
        \hline \\
         0.02643 &0.02860 & 0.324&0.02887&0.340&0.02892\\
         0.02679&0.02860   & 0.330 &0.02860&0.346&0.02893\\
         0.02716& 0.02860  & 0.350 &0.02892&0.339&0.02890 \\
         0.02752& 0.02860  & 0.347&0.02892&0.355&0.02894\\
         0.02643 &   0.02878  &0.339&0.02892&0.354&0.02896\\ 
         0.02679 &   0.02878  & 0.346 &0.02893&0.361&0.02896\\
         0.02716&   0.02878  &0.361&0.02895&0.360&0.02895 \\
         0.02752&   0.02878  &0.364&0.02896&0.375&0.02897 \\ 
    \end{tabular}
    \caption{The parameters $\beta$ and $T_c$ for the U(1) and the U(1) gauge lattice of the parameter set one fitted for eight different temperature intervals via the minima of the least square error. This alternative approach give us $\beta_{\text{U(1) gauge}}=0.345 \pm 0.013$,
    $\Tilde{T}_{c,\text{U(1) gauge}}=0.02888 \pm 0.0001$,  $\beta_{\text{BEC}}=0.354 \pm 0.11$ and $\Tilde{T}_{c,\text{BEC}}= 0.02894\pm 0.00002$.
    }
    \label{tab:beta_power0_LS}
\end{table}

\begin{figure}
    \centering
    \begin{overpic}[width=0.5\linewidth]{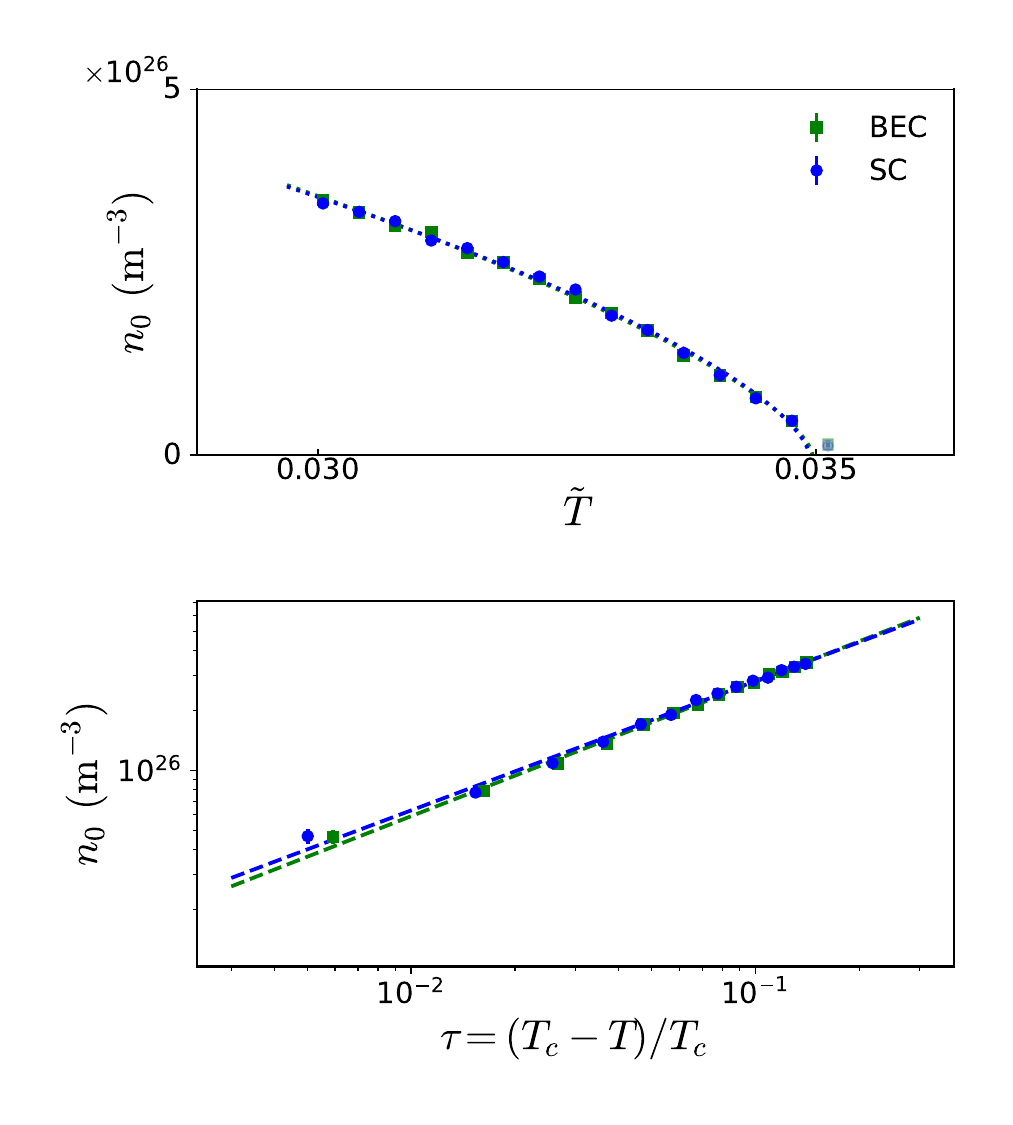}
    \put(0,93){(a)}
    \put(0,43){(b)}
    \end{overpic}
    \caption{
    (a) The Cooper pair density $n_0$ of parameter set two as a function of unitless temperature $\Tilde{T}$  for the superconducting case in  the U(1) gauge theory (SC, blue)  and  the condensate density of a Bose-Einstein condensate without gauge field (BEC, green) for comparison. The shown fit is from the temperature interval $\Tilde{T} \in [0.03005,0.03476]$. \\
    (b) Logarithmic plot of the Cooper pair density $n_0$ as a function of reduced temperature $\tau=(T_c-T)/T_c$ for the same temperature fitting interval. The average extracted critical exponent of all temperature intervals  for the superconducting case with gauge string, $\beta_{\text{U(1) gauge}}=0.344 \pm 0.015$ fits with the BEC case and the expected value for the U(1) universality class.
    }
    \label{figSM3}
\end{figure}

For completeness is in \autoref{figSM3} the Cooper pair density depicted for the parameter set two with very small fluctuations in the particle number per lattice side. In the plot, we show the fit of the maximal used temperature interval $\Tilde{T} \in[0.03005,0.03476]$. 
The fit values of $\beta$ and $T_c$ are again evaluated for different temperature intervals and listed in \autoref{tab:beta_power4}.
All fits together result for parameter set two in $\beta_{\text{U(1) gauge}}=0.344 \pm 0.15$, $\Tilde{T}_{c,\text{U(1) gauge}}=0.03488 \pm 0.00004$, $\beta_{\text{BEC}}=0.354 \pm 0.12$ and $\Tilde{T}_{c,\text{BEC}}= 0.03501\pm 0.00003$.\\
In addition, we list in \autoref{tab:beta_power4_LS} again the parameters obtained via the minimization of the least square error, which are in the standard deviation of the parameters obtained from the direct fitting approach.\\

\begin{table}
    \centering
    \begin{tabular}{p{2.0cm} p{2.0cm} p{2.0cm} p{2.0cm}p{2.0cm}p{2.0cm}}
        $T_{\text{start}}$ & $T_{\text{end}}$ & $\beta_{\text{U(1) gauge}}$ & $T_{c, \text{U(1) gauge}}$ & $\beta_{\text{U(1) }}$ & $T_{c, \text{U(1)}}$\\
        \hline \\
         0.03005 & 0.03476 & 0.325 & 0.03493 &0.338 &0.03497 \\
         0.03041  & 0.03476 & 0.335 & 0.03496  & 0.344 & 0.03498\\
         0.03078& 0.03476 & 0.344 & 0.03498 & 0.353& 0.03501\\
         0.03114 & 0.03476 & 0.345 & 0.03498 & 0.370 & 0.03505\\
         0.03150 & 0.03476 & 0.370 & 0.03504 & 0.364&0.03503 \\
    \end{tabular}
    \caption{The parameters $\beta$ and $T_c$ for the U(1) and the U(1) gauge lattice of the parameter set two fitted for five different temperature intervals. All fits together result for parameter set two in $\beta_{\text{U(1) gauge}}=0.344 \pm 0.15$, $\Tilde{T}_{c,\text{U(1) gauge}}=0.03488 \pm 0.00004$, $\beta_{\text{BEC}}=0.354 \pm 0.12$ and $\Tilde{T}_{c,\text{BEC}}= 0.03501\pm 0.00003$.}
    \label{tab:beta_power4}
\end{table}

\begin{table}
    \centering
    \begin{tabular}{p{2.0cm} p{2.0cm} p{2.0cm} p{2.0cm}p{2.0cm}p{2.0cm}}
        $T_{\text{start}}$ & $T_{\text{end}}$ & $\beta_{\text{U(1) gauge}}$ & $T_{c, \text{U(1) gauge}}$ & $\beta_{\text{U(1) }}$ & $T_{c, \text{U(1)}}$\\
        \hline \\
         0.03005 & 0.03476 & 0.331&0.03496&0.342&0.03500 \\
         0.03041  & 0.03476 & 0.341&0.03498&0.349&0.03501\\
         0.03078& 0.03476 & 0.349&0.03500&0.358&0.03503\\
         0.03114 & 0.03476 &0.351&0.03500&0.371&0.03507\\
         0.03150 & 0.03476 &0.375&0.03505&0.373&0.03505 \\
    \end{tabular}
    \caption{The parameters $\beta$ and $T_c$ for the U(1) and the U(1) gauge lattice of the parameter set two fitted for five different temperature intervals via the minimization of the least square error. We obtain $\beta_{\text{U(1) gauge}}=0.349 \pm 0.015$, $\Tilde{T}_{c,\text{U(1) gauge}}=0.03500 \pm 0.00002$, $\beta_{\text{BEC}}=0.359 \pm 0.12$ and  $\Tilde{T}_{c,\text{BEC}}= 0.03503\pm 0.00002$}
    \label{tab:beta_power4_LS}
\end{table}

\subsection{Vortices} \label{subsec:vortices}
\begin{figure}
    \centering
    \includegraphics[width=0.5\linewidth]{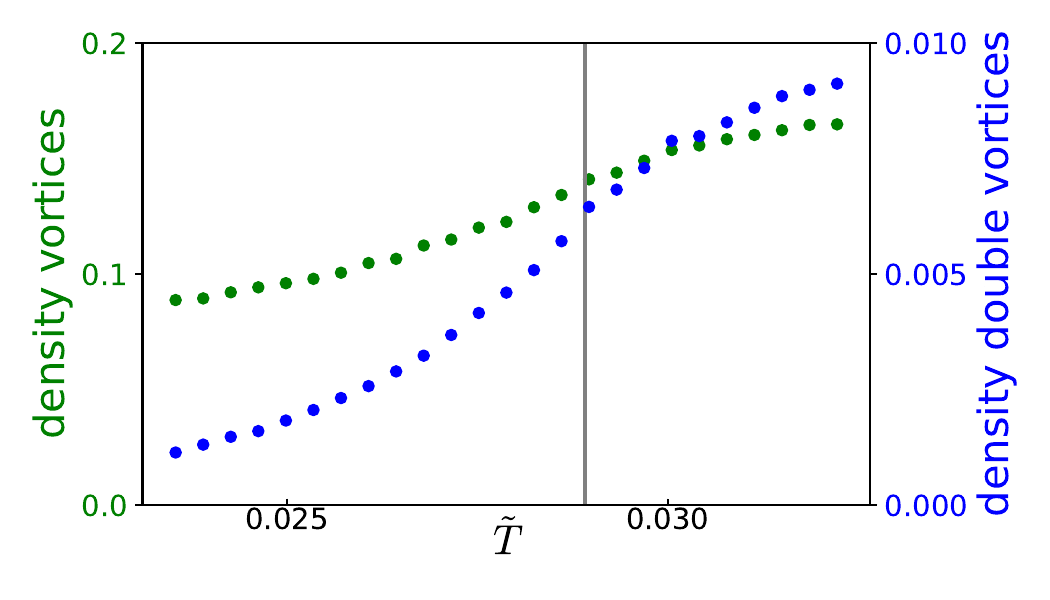}
    \caption{The average density of vortices and double vortices per lattice plaquette as a function of unitless temperature $\Tilde{T}$ for the $14^3$ lattice.}
    \label{figSM4}
\end{figure}

In \autoref{figSM4} is the  density of the vortices per plaquette over temperature depicted in green (left axis) for parameter set one. The density of vortices is increasing with temperature, thereby the highest slope is reached at the critical temperature (grey line).
Plotted in blue (right axis) is the density of double vortices with $\nu =\pm 2$. Here the same behavior of increase and maximum slope at $\Tilde{T_c}$ can be observed.\\

We further analyzed the spatial correlation of the vortices in the xy-plane of our isotropic system, using the vortex correlation function \cite{Homann2024}
\begin{align}
    V^{xy}(x_i,y_j)=\frac{\expval{\nu_{l,m,n}^{xy}*\nu_{l+i,m+j,n}^{xy}}}{\expval{\nu_{l,m,n}^{xy}}^2}.
\end{align}

\begin{figure}
    \centering
    \begin{overpic}[width=0.8\linewidth]{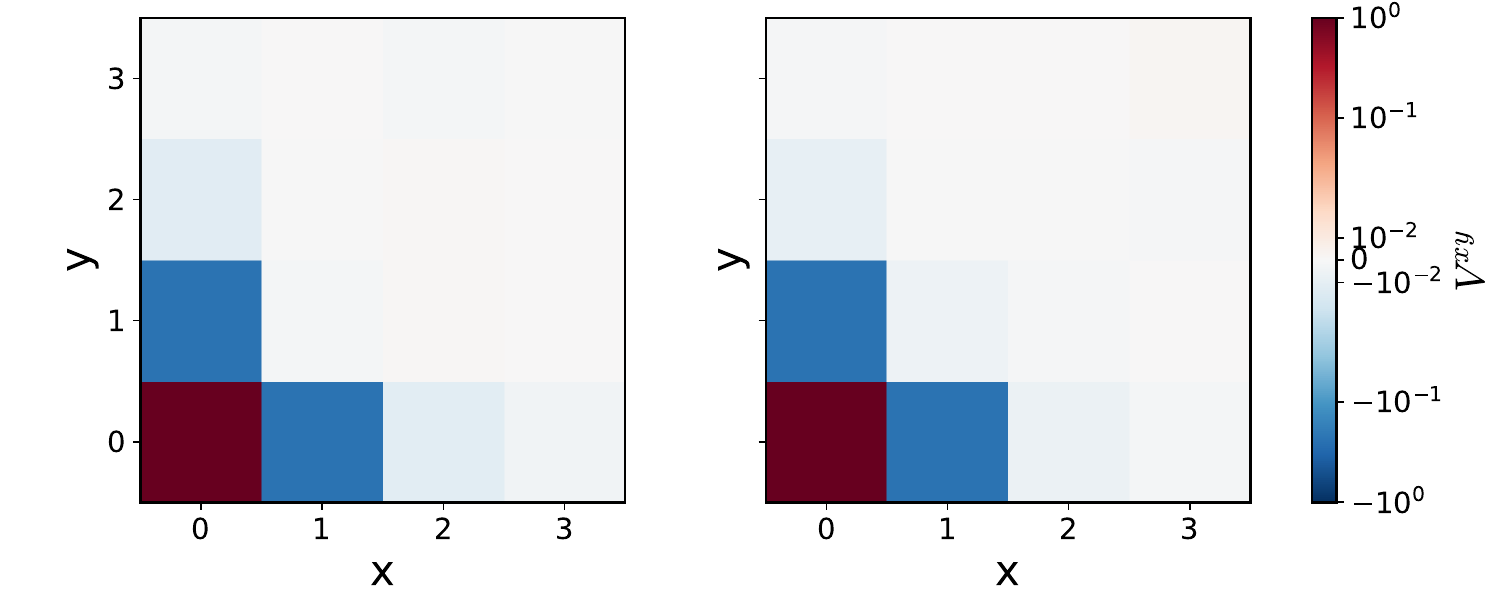}
    \put(2,39){(a)}
    \put(46,39){(b)}
    \end{overpic}
    \caption{Vortex correlation function at (a) $\Tilde{T}=0.0235$ below the critical temperature and (b) $\Tilde{T}=0.0326$ above the critical temperature. There are high correlations to find a vortex of opposite vorticity on the nearest-neighbour plaquette for both temperatures. For $\Tilde{T}=0.0326$, above the critical temperature, a higher correlation for the diagonal element can be observed. This can be explained with larger vortex clusters and the higher vortex density for higher temperatures.}
    \label{figSM5}
\end{figure}

\begin{figure}
    \centering
    \includegraphics[width=0.5\linewidth]{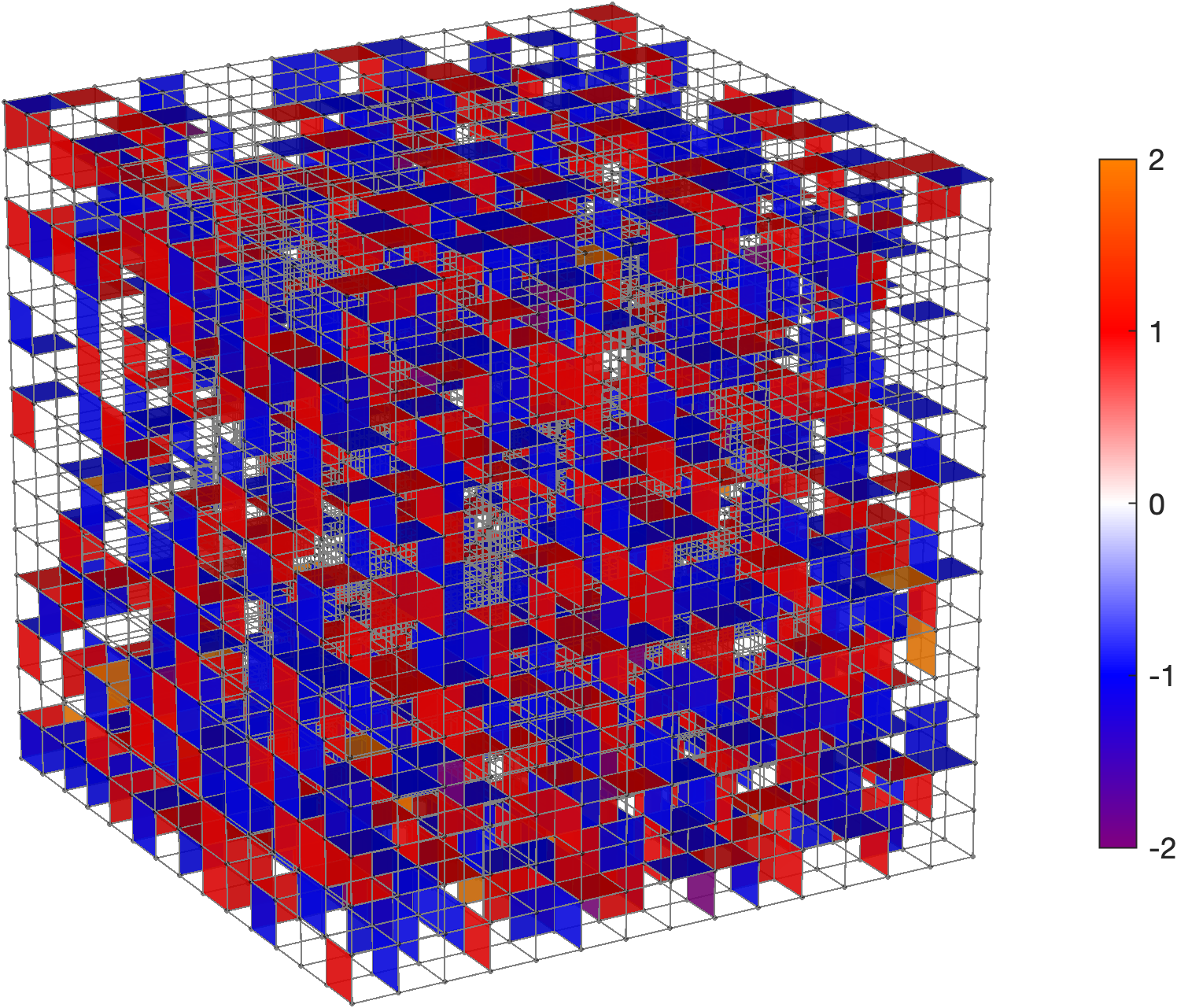}
    \caption{Screenshot for the vorticity on the $14^3$ lattice for $\Tilde{T}=0.0272$ and parameter set one. }
    \label{figSM6}
\end{figure}

In \autoref{figSM5} the vortex correlation function  for the temperatures $\Tilde{T}=0.0235$ below and $\Tilde{T}=0.0326$ above the critical temperature for parameter set one are shown.
We see the expected high correlation of finding a vortex of opposite vorticity next to a vortex for both temperatures.
Further there is a  small increase of correlation on the diagonal next nearest neighbor plaquette from \autoref{figSM5}~(a) to \autoref{figSM5}~(b), corresponding to  larger clusters and  larger vortex density with increasing temperature. The vortex density and the spatial vortex correlation function are evaluated for the $14^3$ lattices. Last, depicted in \autoref{figSM6} is a screenshot of the vorticity on the $14^3$ lattice for $\Tilde{T}=0.0272$ and parameter set one.\\

\subsection{Anomalous dimension} \label{subsec:ana_dim}

\begin{figure}
    \centering
    \begin{overpic}[width=0.3\linewidth]{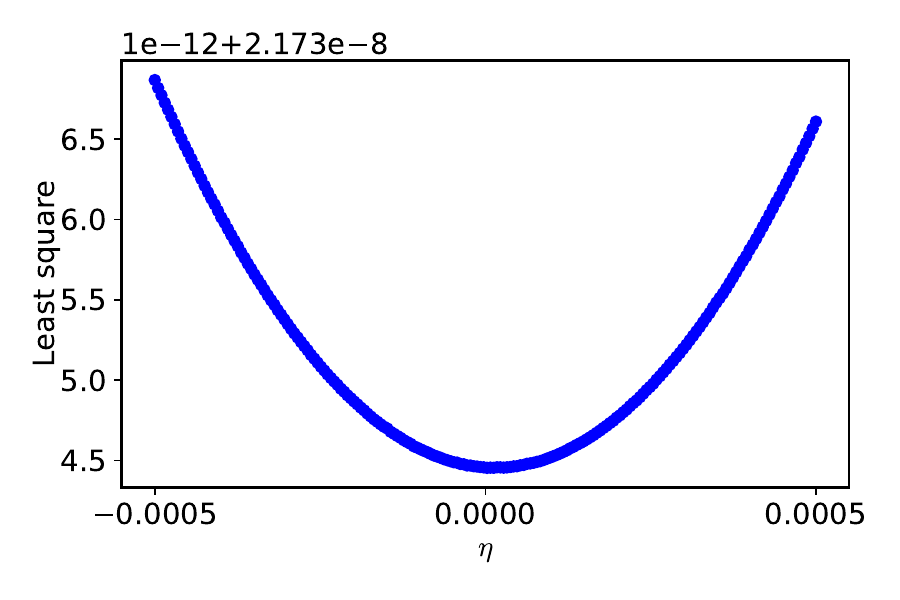}
    \put(-2,58){(a)}
    \end{overpic}
    \centering
    \begin{overpic}[width=0.3\linewidth]{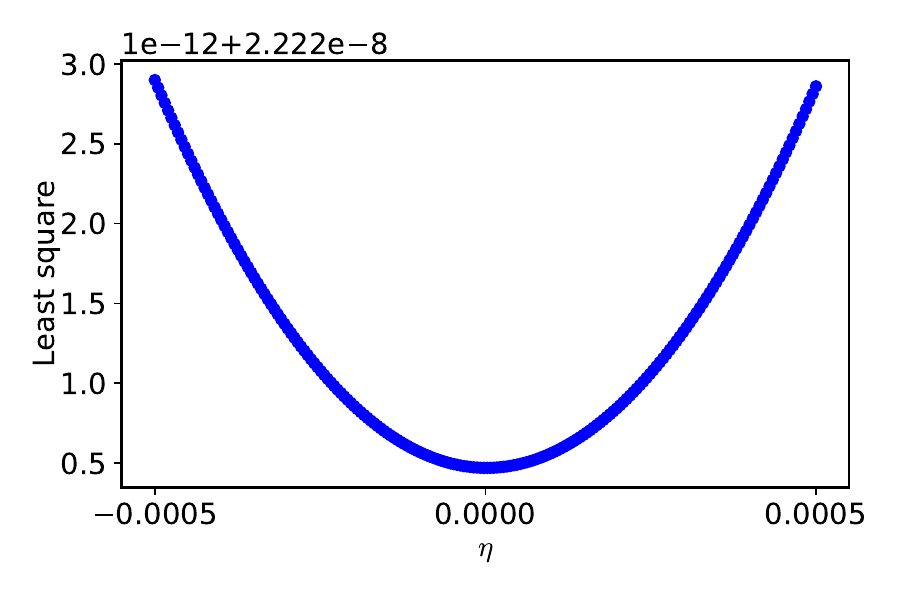}
    \put(-2,58){(b)}
    \end{overpic}
    \centering
    \begin{overpic}[width=0.3\linewidth]{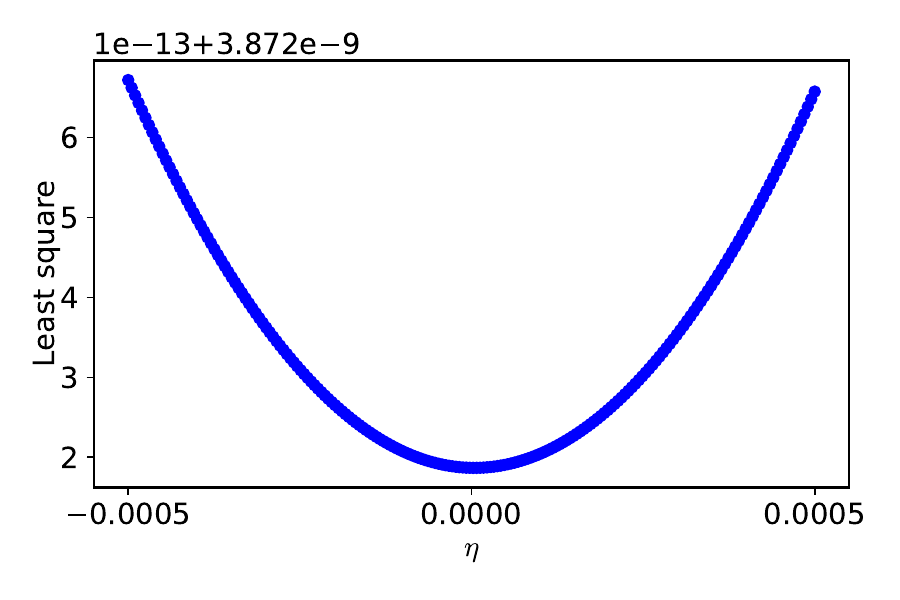}
    \put(-2,58){(c)}
    \end{overpic}
    \caption{Least square error for the fit of the cumulated correlation function for additional temperatures and the second parameter set. For all parameter combinations the best $\eta$ is in the range of $(-1,1)\times10^{-4}$. (a) $\Tilde{T}=0.0253$ and parameter set 1. (b)  $\Tilde{T}=0.0308$ and parameter set 2. (c)  $\Tilde{T}=0.0344$ and parameter set 2.}
    \label{figSM7}
\end{figure}

We show in \autoref{figSM7} the least square error as a function of $\eta$ for an additional temperature $\Tilde{T}=0.0253$ of parameter set 1 (\autoref{figSM7}~(a)) and for two temperatures of  parameter set 2 (\autoref{figSM7}~(b) $\Tilde{T}=0.0308$, \autoref{figSM7}~(c) $\Tilde{T}=0.0344$). For all parameter combinations we can determine the best $\eta$ in the range of $(-1,1)\times10^{-4}$.

\end{document}